\documentclass{ws-p9-75x6-50}

\begin{document}

\def\beq{\begin{equation}}
\def\eeq{\end{equation}}
\def\d{{\delta}}

\title{Superfluid neutron stars}

\author{David Langlois}

\address{ Institut d'Astrophysique de Paris, \\
Centre National de la Recherche Scientifique,\\
98bis Boulevard Arago, 75014 Paris, France \\
and\\
D\'epartement d'Astrophysique Relativiste et de Cosmologie,\\
Centre National de la Recherche Scientifique,\\
Observatoire de Paris, 92195 Meudon Cedex, France}


\maketitle

\abstracts{Neutron stars are believed to contain  (neutron and proton) superfluids. I will give a summary of a macroscopic  description of the interior of neutron stars, in a  formulation which is general relativistic. I will also present recent results on the oscillations of neutron stars, with superfluidity explicitly taken into account, which leads in particular to the existence of a new class of modes.			
 }

\section{Superfluid relativistic hydrodynamics}
Neutrons stars are believed to contain in their interior superfluid neutrons
and superconducting protons. This is suggested by nuclear physics calculations
and by the very long relaxation time scales after `glitches' (sudden 
increases of the neutron star angular velocity).
Over the last years, we have been developing a formalism that can describe 
the relativistic hydrodynamics of mixtures of superfluid or superconductors
and possibly ordinary fluids \cite{cl,lsc98,cl98}.
\font\srm=cmr9
\def\eX{e^{_{\rm X}}}
\def\piX{\pi^{_{\rm X}}}
\def\wX{w^{_{\rm X}}}
\def\nX{n_{_{\rm X}}}
\def\muX{\mu^{_{\rm X}}}
 
The hydrodynamics of  several species labelled by the index {\srm X}, with 
respective particle currents $\nX^{\,\rho}$ and respective 
electric charge per particle $\eX$, can be described by  
the  Lagrangian
\beq
{\cal L}=\Lambda_{\rm M}+\sum_{\rm X}\eX \nX^{\,\rho} A_\rho+{1\over 16\pi} 
F_{\rho\sigma}F^{\sigma\rho}\, ,
\eeq
where $A_\rho$ is the electromagnetic gauge form and $F_{\rho\sigma}$ 
the corresponding electromagnetic tensor, and where $\Lambda_{\rm M}$ 
is the `hydrodynamical' part of the Lagrangian, depending only on 
the particle currents $\nX^{\,\rho}$, the variations of which 
define the momentum covectors: $\delta \Lambda_{\rm M}=\muX_{\ \rho}
\delta \nX^{\,\rho}$. This variational principle, corresponding to 
perfectly conducting fluids, leads to  
 matter 
conservation equations, one for each species,
$\nabla_\rho \,\nX^{\,\rho}=0$  (one can generalize to allow for 
chemical reactions between various species \cite{lsc98}),  and to 
 Euler-type equations of motion, which can be written in the very compact form
\beq
\nX^{\,\sigma}\nabla_{[\rho}\piX_{\sigma]}=0, \quad {\rm with} \quad 
\piX_{\ \rho}\equiv\muX_{\ \rho}+\eX A_\rho\, 
\eeq

In order to deal with  superfluids or superconductors,  one must impose
the further condition that the momentum covectors 
$\piX_{\ \rho}$ are gradients. This is in fact 
true on small scales only, because on larger scales the superfluid or the 
superconductor will be threaded in general by arrays of vortices, due 
to a global angular momentum (for superfluids) or an external magnetic field
(for superconductors).
The macroscopic description, which takes into account the average effect 
of the vortices, can be 
done by considering the (generalized) vorticity tensors 
$\wX_{\rho\sigma}\equiv 2\nabla_{[\rho}\piX_{\sigma]}$, now non-zero 
since there are vorticies, as  fundamental quantities \cite{cl98}.

\section{Oscillations of superfluid neutron stars}
In a neutron star, the existence of a superfluid component, weakly 
connected to the normal component, leads to a richer spectrum of 
oscillations. In general relativity, one must take into 
account not only the matter perturbations but also the perturbations
of the metric $g_{\mu\nu}$.
In a specific gauge, the even-parity 
modes  can be written (the study of the  $m=0$ modes is enough because 
of the degeneracy in $m$)  in  the 
form 
\begin{eqnarray}
&\delta g_{00}=-e^\nu r^l H_0 e^{i\omega t}P_l(\theta), 
&\delta g_{0r}=\delta g_{r0}= -i\omega r^{l+1} H_1 e^{i\omega t}P_l(\theta),\cr
&\delta g_{rr}= -e^\lambda r^l H_2 e^{i\omega t}P_l(\theta), 
&\delta g_{\theta \theta}=\delta g_{\phi\phi}/\sin^2\theta=
- r^{l+2} K e^{i\omega t}P_l(\theta),
\end{eqnarray}
where $P_l$ stands for the Legendre polynomial of order $l$.
In a two-component model, the matter perturbations can
 be described by the two 
matter displacements $\xi_n$ and $\xi_p$, 
\begin{eqnarray}
\xi_n^r=&r^{l-1}e^{-\lambda/2}W_n e^{i\omega t}P_l(\theta), \quad
\xi_n^\theta=&- r^{l-2}V_n e^{i\omega t}\partial_\theta P_l(\theta),\cr
 \xi_p^r=&r^{l-1}e^{-\lambda/2}W_p e^{i\omega t}P_l(\theta), \quad
\xi_p^\theta=&- r^{l-2}V_p e^{i\omega t}\partial_\theta P_l(\theta).
\end{eqnarray} 
Inserting the above expressions for the perturbations into the perturbed 
Einstein's equations as well as the perturbed Euler equations, one ends 
up with a system of linear equations, consisting of two contraints (one 
is $H_2=H_0$, the other  expresses  $H_0$ in terms of the other 
perturbations) and of first order differential equations of the form 
\beq
{dY\over dr}= Q_{l,\omega}Y, 
\eeq
where $Y$ is 6-dimensional column vector containing $H_1$, $K$, $W_n$, $V_n$, 
$W_p$ and $V_p$, and $Q_{l,\omega}$ is a $6\times 6$ matrix with
$r$-dependent  coefficients
that depend only on the background configuration, as well as on $l$ and 
$\omega$.
Considering the boundary conditions at the center and at the surface of the 
star, this system can be solved, up to a global amplitude, for {\it any value} 
of $\omega$. The physically relevant modes, however, also called quasi-normal 
modes, correspond  the specific values 
of $\omega$ for which the metric outside 
the star represents {\it only outgoing gravitational waves}.

A numerical investigation for a very crude  model of two 
independent polytropes (with different adiabatic indices) has shown that 
the two-component star will exhibit new modes, {\it superfluid} modes, 
which are specific of the existence of two components since the two fluids 
are counter-moving for these modes \cite{cll}.


\begin{thebibliography}{99}


\bibitem{cl}  B. Carter, D. Langlois, {\it Phys. Rev.} D 51, 5855 (1995);
  {\it Phys. Rev.} D 52, 4640 (1995); 
{\it Nuclear Physics} B 454, 402 (1995).

\bibitem{lsc98} D. Langlois, D. M. Sedrakian,  B. Carter,  {\it M.N.R.A.S.}
297, 1189 (1998).

\bibitem{cl98} B. Carter, D. Langlois, {\it Nuclear Physics}  {\bf B 531}
, 478 (1998)

\bibitem{cll} G.L. Comer, D. Langlois, Lap Ming Lin, {\it Phys. Rev.} {\bf D 60}, 104025 (1999)

\end{thebibliography}
\end{document}